\documentclass[sigconf, screen]{acmart}
\usepackage{graphicx}

\AtBeginDocument{%
  \providecommand\BibTeX{{%
    \normalfont B\kern-0.5em{\scshape i\kern-0.25em b}\kern-0.8em\TeX}}}

\copyrightyear{2024}
\acmYear{2024}
\setcopyright{acmlicensed}\acmConference[LAK '24]{The 14th Learning Analytics and Knowledge Conference}{March 18--22, 2024}{Kyoto, Japan}
\acmBooktitle{The 14th Learning Analytics and Knowledge Conference (LAK '24), March 18--22, 2024, Kyoto, Japan}
\acmPrice{15.00}
\acmDOI{10.1145/3636555.3636858}
\acmISBN{979-8-4007-1618-8/24/03}

\def\sep{\, | \,}

\begin{document}

\title{Adaptation of the Multi-Concept Multivariate Elo Rating System to Medical Students' Training Data}

\author{Erva Nihan Kandemir}
\email{erva.nihan.kandemir@ens.psl.eu}
\affiliation{
  \institution{École Normale Supérieure, PSL University, CNRS}
  \city{Paris}
  \state{75005}
  \country{France}
}

\author{Jill-Jênn Vie}
\email{vie@jill-jenn.net}
\affiliation{
  \institution{Soda team, Inria Saclay}
  \country{France}
}

\author{Adam Sanchez-Ayte}
\email{adam.sanchez@uness.fr}
\affiliation{
  \institution{Université Numerique en Santé et Sport (UNESS)}
  \country{France}
}

\author{Olivier Palombi}
\email{olivier.palombi@univ-grenoble-alpes.fr}
\affiliation{
  \institution{Université Grenoble Alpes, Grenoble INP, CNRS, Inria, LIG}
  \streetaddress{P.O. Box 1212}
  \city{Grenoble}
  \state{38000}
  \country{France}
}

\author{Franck Ramus}
\email{franck.ramus@ens.psl.eu}
\affiliation{
  \institution{École Normale Supérieure, PSL University, EHESS, CNRS}
  \city{Paris}
  \state{75005}
  \country{France}
}

\renewcommand{\shortauthors}{Kandemir, Vie, Sanchez-Ayte, Palombi, and Ramus}

\begin{abstract}
Accurate estimation of question difficulty and prediction of student performance play key roles in optimizing educational instruction and enhancing learning outcomes within digital learning platforms. The Elo rating system is widely recognized for its proficiency in predicting student performance by estimating both question difficulty and student ability while providing computational efficiency and real-time adaptivity. This paper presents an adaptation of a multi-concept variant of the Elo rating system to the data collected by a medical training platform—a platform characterized by a vast knowledge corpus, substantial inter-concept overlap, a huge question bank with significant sparsity in user-question interactions, and a highly diverse user population, presenting unique challenges. Our study is driven by two primary objectives: firstly, to comprehensively evaluate the Elo rating system's capabilities on this real-life data, and secondly, to tackle the issue of imprecise early-stage estimations when implementing the Elo rating system for online assessments. Our findings suggest that the Elo rating system exhibits comparable accuracy to the well-established logistic regression model in predicting final exam outcomes for users within our digital platform. Furthermore, results underscore that initializing Elo rating estimates with historical data remarkably reduces errors and enhances prediction accuracy, especially during the initial phases of student interactions. 

\end{abstract}





\sloppy
\maketitle

\section{Introduction}
Over the last decade, as a result of the increasing use of educational technology involving significant amounts of data and learning analytics, personalized learning has gained increasing popularity. This has led a number of research groups to study the adaptation of personalized learning into educational technologies, leveraging insights from learning analytics \cite{lee2020learning, aljawarneh2021data}. 
Adaptive Learning Systems (ALSs) \cite{lee2008adaptive}
achieve personalized learning experiences by using users' prior interactions and by adjusting the learning content to match individual preferences and requirements. Research consistently highlights the effectiveness of ALSs when compared to non-adaptive systems, resulting in improved learning outcomes \cite{lindsey2014improving, tabibian2019enhancing, vanlehn2011relative, ma2014intelligent} and a positive impact on student motivation \cite{abdi2019multivariate}, engagement \cite{papousek2015impact}, and comprehension \cite{alamri2020using}. Today, prominent ALS platforms like Duolingo, and ALEKS deliver high-quality learning materials and personalized instruction to millions of users worldwide.

With all the benefits listed above, this adaptive method in online education requires effectively monitoring users' learning paths, a procedure called knowledge tracing. This knowledge-tracing process involves creating learner models based on student performance and interactions with the system to represent their ability levels and the difficulty of educational materials.

\subsection{Background}

Various models have been designed to monitor and predict students' evolving knowledge levels over time \cite{brooks2017predictive}. These models can be broadly classified into four categories: Markov process models \cite{corbett1994knowledge,gweon2015tracking}, logistic models \cite{choffin2019das3h,rasch1960studies,vanderlinden2013handbook,lindsey2014improving}, deep knowledge tracing models \cite{piech2015deep,ghosh2020context,chan2021clickstream,ruan2021variational,shin2021saint}, and rating systems \cite{abdi2019multivariate}. 
The first three classes of these models are well-established and already extensively documented in existing literature \cite{abdelrahman2023knowledge}. Although they exhibit strong prediction capabilities when evaluating student performance, they are not without limitations, particularly when deploying them in online educational environments, where they often demand intricate parameter estimation and calibration procedures, typically relying on large datasets. This complexity can impede the development of adaptive systems, making them more challenging, time-consuming, and resource-intensive to create. 

An intriguing alternative is the utilization of rating systems, offering a computationally more economical approach. Rating Systems, particularly the Elo Rating system \cite{elo1978rating}, have been widely applied in educational technologies \cite{attali2014ranking, pelanek2016applications, papousek2017should, klinkenberg2011computer}. Originally created for ranking chess players, the Elo rating system has been repurposed for educational settings, treating users and learning materials as opponents. In the educational context, it predicts ratings for users and questions, serving as an assessment of user ability and question difficulty.

In this framework, each user $u$ is associated with a global ability parameter denoted as $\theta_u$. Similarly, for each question $i$, there exists a question parameter $\theta_i$ reflecting the difficulty level of that question. The probability of a user $u$ correctly attempting a multiple-choice question $i$, denoted as $\Pr(a_{ui}=1 | \theta_u, \theta_i)$, can be expressed as a logistic function of the difference between the user and question parameters:
\begin{displaymath}
 \ \Pr(a_{ui}=1 | \theta_u, \theta_i ) = \sigma(\theta_{u} - \theta_i) = \frac{1}{1 + e^{-(\theta_{u} - \theta_i)}} 
\end{displaymath}

After a user $u$ attempts question $i$, both the user's ability and the question's difficulty undergo updates that are proportional to the difference between the estimated probability and the actual outcome. These updates are defined by the following formulas for question difficulty (\(\theta_i\)) and for user ability (\(\theta_u\)):
\begin{equation}
\begin{aligned}
\label{elo-update-rule}
\theta_i :=  \theta_i + K (\Pr(a_{ui}=1|\theta_{u}, \theta_i) - a_{ui})\\
 \theta_{u} := \theta_{u} + K (a_{ui} - \Pr(a_{ui}=1|\theta_{u}, \theta_i)) 
 \end{aligned}
\end{equation}

Here, \(a_{ui}\) represents the actual outcome of the attempt of the user \(u\) on question \(i\), and \(K\) is a constant value that determines the degree of update sensitivity based on the user's most recent attempt. The choice of the constant $K$ in the update rule plays a pivotal role in shaping estimation dynamics. If $K$ is set too low, the estimation process progresses too slowly, leading to prolonged uncertainty in skill assessment and potential failure to reach correct values. Conversely, if $K$ is set too high, the system is unstable, heavily influenced by recent attempts, and thus provides erratic evaluations.

In light of this formulation, the classical Elo rating system in education reveals its iterative nature, refining user and question parameters after each interaction. 

While the Elo rating system does not provide statistically guaranteed estimations, in contrast to well-calibrated logistic models, numerous studies have explored the accuracy of the Elo rating system and compared its performance to state-of-the-art models. For instance,  study \cite{wauters2012item} found that the IRT-Rasch model version, proportion correct, and the Elo rating system, increasingly correlate with the true difficulty parameter as sample size increases. Another study \cite{papousek2014adaptive} compared Elo rating's question difficulty estimates with those obtained through the joint maximum likelihood method (JMLE) and observed nearly identical outcomes. Studies using simulated data \cite{antal2013elo, pelanek2014application, pelanek2016applications} have also concluded that Elo rating systems perform similarly to the Rasch model, making them suitable for systems requiring real-time user knowledge adaptation without the need for extensive question pretesting on large sample sizes.

However, integrating the Elo rating algorithm into a real-time educational context presents challenges, especially in the initial stages when student abilities and question difficulties are unknown and are set to zero. Given the iterative nature of the model, these initial estimates are assumed to gradually correct themselves with each attempt. While starting the model from scratch is standard practice, to produce reliable estimates the system requires a substantial number of responses, typically at least 100 attempts \cite{pelanek2016applications}. 
Furthermore, uncertain initial estimates can exert a lasting influence on subsequent updates, potentially resulting in persistent inaccuracies. This issue is especially pronounced for questions and users with a limited number of attempts within the educational platform, as they may have fewer opportunities for correction.


\subsection{Goals of the present study}

In this study, we have two primary objectives: firstly, to assess whether the Elo rating system meets the requirements of a complex real-life scenario with specific challenges, and secondly, to address the issue of imprecise early-stage estimations in the online application of the Elo rating system. To mitigate the uncertainty associated with initial Elo rating estimations, we used data collected in the previous year.

In brief, our learning platform is open to all French medical students throughout their studies to provide training for their medical studies (about 8600 students per year over 10 years, with one important national exam at the end of the 6\textsuperscript{th} year). The challenges raised by this particular learning context are multiple:

\begin{itemize}
\item The knowledge corpus is huge, as it encompasses \textit{all medical knowledge} taught in French universities.
\item This corpus is structured into knowledge components that are themselves very large: 362 distinct topics or subcategories (themes/areas) of medical knowledge, spread over 31 medical specialties.
\item The corpus of questions is also huge (\(\sim\)1,500,000 questions), such that a given student takes only a tiny fraction of available questions (usually drawn at random), almost never takes the same question twice, and such that a given question is only taken by a tiny fraction of students (outside exams). Thus, the matrix is extremely large and sparse.
\item Use of the training platform is optional, with some students using it intensively on a daily basis, and others doing most of their training outside the platform.
\item Students are based in 42 universities which cover the same curriculum from 1\textsuperscript{st} to 6\textsuperscript{th} year, but with different material and in a different order.
\end{itemize}

Despite these difficulties, the fact that all medical students from all universities can train and take exams on the same platform should make it possible and desirable to model their progress and use this modeling to provide them with an adaptive training program. Yet the first step is to show that a sufficiently reliable modeling of their progress is possible under the present conditions.

Thus, the main research question guiding this study is to examine the boundaries of the Elo rating system within a particularly challenging real-life scenario that encompasses various complexities. This evaluation involves a comparison of its accuracy against the widely accepted logistic regression model. Additionally, we seek to explore the potential advantages of initializing the Elo system with results obtained from logistic regression applied to data from preceding years. 

\section{METHOD}

This study employed an observational research design, leveraging ecological data from the existing BNE (\emph{Banque nationale d'entraînement}) digital learning platform.

\subsection{BNE Platform}
The BNE digital learning system serves as an online platform extensively utilized by over 8,800 medical students across all French universities annually. This platform is used by all medical faculties to administer exams. Exam questions are then added to the question bank, together with additional questions designed by professors for training purposes. This platform therefore holds a very large set of multiple-choice questions covering 31 medical specialties that are made available to students for training. For medical students, the platform is a valuable resource to prepare for the ECN (\emph{Épreuves Classantes Nationales}) final exam. This final exam usually takes place in June of the sixth academic year and significantly influences the choice of students' medical specialization.

To enhance the pedagogical engagement of medical students on the platform, a notable feature allows learners to tailor their training experience. They can choose from various question types and medical specialties, enabling them to simulate and practice for a wide range of medical exams according to their preferences and needs.

\subsection{BNE data set Overview}

Within this section, we describe the BNE data set for the educational year 2020-2021, representing the most recent accessible data sourced from the BNE platform during our analysis. Additionally, we describe the usage patterns observed on the platform during this year, providing valuable insights into its structure and functionality. 

From the BNE platform, direct access to the official ECN exam is unavailable. However, we do have access to a mock exam, typically conducted in mid-March (specifically on March 15\textsuperscript{th}, 16\textsuperscript{th}, and 17\textsuperscript{th}, 2021, for the 2020-2021 academic year). This mock exam closely mimics the format of the actual ECN final exam. For the purpose of testing student prediction models on our complex data set, we focused our analysis on the educational year of 2020-2021, specifically targeting 6\textsuperscript{th}-year users who participated in the mock final ECN exam. This selection aligns with the core objective of our study, which is to assess the models' performance through external validation using the mock final exam.

In the following sections, we will describe the data related to the six-month training period spanning from September 16, 2020 to March 14, 2021, leading up to the mock final exam, distinct from the data associated with the mock exam itself, on March 15\textsuperscript{th}, 16\textsuperscript{th}, and 17\textsuperscript{th}, 2021.

\begin{table*}
  \centering
  \caption{\textbf{BNE data set Summary}}
  \label{tab:summary}
  \begin{tabular}{cc|ccccccccc}
    \toprule
    Data &  Period & \rotatebox{90}{Users} & \rotatebox{90}{Questions} & \rotatebox{90}{Specialties} & \rotatebox{90}{Specialties per Question} & \rotatebox{90}{Attempts} & \rotatebox{90}{Sparsity (User, Question)} & \rotatebox{90}{Attempts per User} \\
    \midrule
    \textbf{Training Period data set} & \textbf{16.09.2020--14.03.2021} & 8,616 & 357,317 & 31 & 1.58 & 26,772,424 & 0.99 & 1.05 \\
    \textbf{Mock Final Exam data set} & \textbf{15--17.03.2021} & 8,616 & 372 & 28 & 1.71 & 3,172,546 & 0.01 & 1 \\
    \bottomrule
  \end{tabular}
\end{table*}

\subsubsection{Training Period data set}

Table~\ref{tab:summary} offers a comprehensive overview of our training period data set's key characteristics, including the total number of \emph{users} (8,616), \emph{questions} (357,317), \emph{medical specialties} (31), and \emph{attempts} (26,772,424). 

\begin{figure}
  \centering
  \includegraphics[width=1\linewidth]{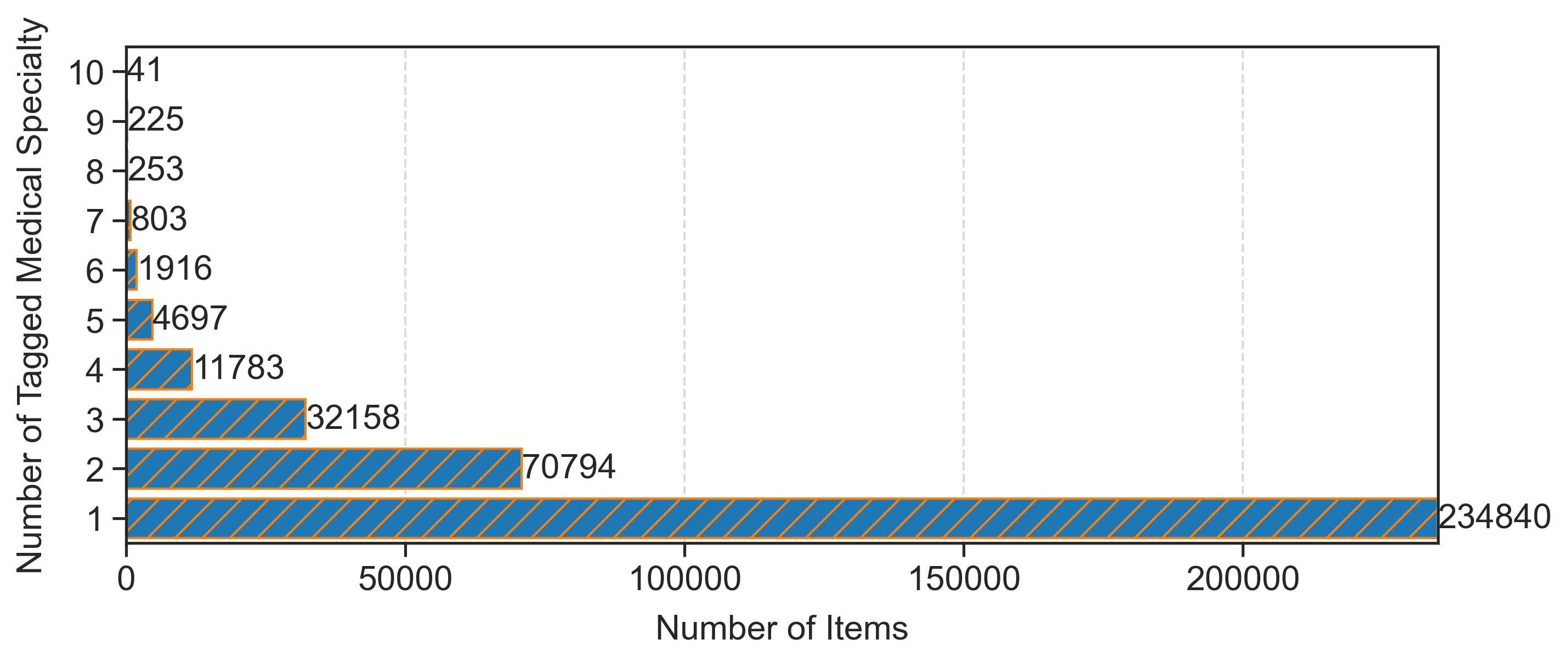}
  \caption{Number of questions that require knowledge on any given number of medical specialties. \\ The questions exhibit a spectrum of dependence on medical specialty knowledge for their solution. While a considerable portion of questions rely on ability in a single medical specialty, many questions require knowledge spanning multiple specialties.}
  \label{fig:question-spectrum}
\end{figure}

Additionally, the \emph{specialty per question} variable indicates the average number of specialty tags associated with each question. Here, each of the 31 medical specialties serves as a distinct knowledge component (KC). These knowledge components are much larger than is usually defined in the literature. 
In this context, when a question $i$ is tagged with a specific specialty $s$, the likelihood of correctly answering question $i$ hinges upon the user's specialty-specific ability $s$. Conversely, we assess a user's ability in specialty $s$ by evaluating their ability to successfully tackle questions tagged with the same specialty $s$. The table reveals that the average number of specialties associated with each question (1.58) is greater than 1, underscoring that certain questions require knowledge spanning multiple medical specialties for accurate responses. In this context, our data set can be characterized as a multi-knowledge component data set (or multi-specialty data set in the specific context of BNE). For a more in-depth exploration of this distribution, Figure~\ref{fig:question-spectrum} provides a detailed breakdown of the count of questions requiring varying numbers of specialties. 

The user-question sparsity in Table~\ref{tab:summary} indicates the proportion of missing values in the user-question interaction matrix. The table shows that our data set is substantial and exhibits significant sparsity (\emph{Sparsity (user, question)} = 0.99). There are vastly more available questions than any user can take, and different users will take different questions (usually by random draw). 

Lastly, the \emph{Attempts per User} variable unveils the average number of attempts made by the same user on individual questions. This indicates whether students frequently revisit questions they have previously encountered. With a value of 1.05, it is evident that during the training period, students almost never re-attempt questions they have already attempted. 

\begin{figure*}
  \centering
  \includegraphics[width=0.7\linewidth]{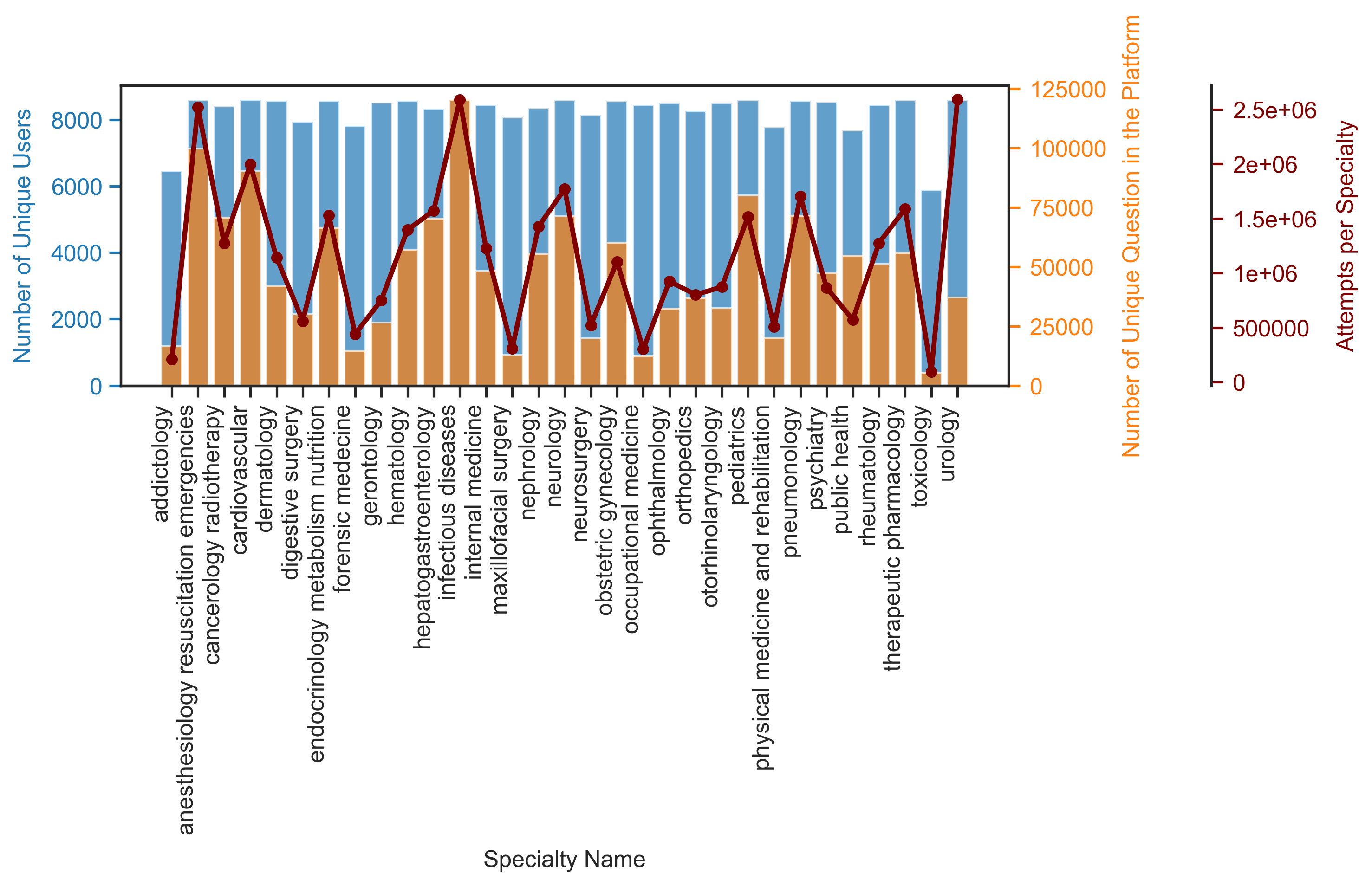}
  \caption{Overview of the use of the BNE Platform during the 2020-2021 educational year. \\
  The blue bars represent the count of unique users per medical specialty in the data set. The orange bars represent the count of unique questions available in the platform for each specialty. In addition, the overlaid line plot illustrates `Attempts per Specialty,' the total number of user attempts on questions within each specialty during the educational year 2020-2021.}
    \label{fig:usageoverview}
\end{figure*}

Figure~\ref{fig:usageoverview} provides a detailed description of the platform's usage patterns across 31 available medical specialties during the 2020-2021 educational year. As mentioned earlier, the platform provides users with the option to create their own training sessions by selecting specific medical specialties and question types they want to study. This results in varying levels of popularity among the specialties. The figure reveals that while most students engage with questions from all specialties, there is considerable variability in both the quantity of available questions and the number of questions taken within each specialty.

\subsubsection{Mock Final Exam data set}

Table~\ref{tab:summary} also provides a comprehensive overview of the nature of the mock ECN final exam for the educational year 2020-2021, held on the 15\textsuperscript{th}, 16\textsuperscript{th}, and 17\textsuperscript{th} of March 2021. The Mock final exam data set encompasses data from 8,616 users who took 372 questions spanning 28 distinct medical specialties (addictology, orthopedics, and toxicology were not included in the mock exam). Questions in this data set were associated with a mean number of 1.71 specialties, reflecting greater multidisciplinarity of questions than in the training period data set. The mock final exam data set recorded a  total of 3,172,546 user interactions, highlighting a high level of user engagement, with a minimal sparsity value of 0.01, implying that almost all users attempted every question during the final exam. Additionally, the \emph{Attempts per User} value of 1 indicates that users made only a single attempt at each question. 

Here, it's worth noting that all the mock exam questions were entirely new, distinct from those encountered during the training period. Therefore,  the data from the mock final exam does not provide a direct test of the knowledge of specific questions taken in the training period, but rather a test of students' ability to generalize what they have learned during courses and training to new questions in the same medical specialties, mirroring the format of the official ECN exam, which emphasizes generalization rather than memorization of specific knowledge.


\subsection{Elo Rating: Model Extensions for Adaptation to the BNE data set}

The standard iterative formulation of the Elo rating system, which computes user and question-related factors, has been previously described in the related literature \cite{attali2014ranking, pelanek2016applications, papousek2017should, klinkenberg2011computer,antal2013elo, pelanek2014application, papousek2014adaptive}. To optimize its adaptability for educational contexts, the Elo rating system has undergone numerous extensions. In this section, we indicate the specific modifications we have applied to tailor the model to our BNE data set.

\subsubsection{Incorporating the guessing behavior into the Elo rating system}

In numerous studies employing the Elo rating as a prediction model for multiple-choice questions, researchers take into account the guessing rate when calculating the probability of correctness \cite{papousek2014adaptive, pelanek2016applications}.

In such instances, the probability of a user $u$ attempting a multiple-choice question $i$ with $n_{\text{opt}}$ choices correctly, denoted as  $\Pr(\text{correct} \, | \, \theta_u, \theta_i)$, can be expressed as follows:
\begin{displaymath}
\ \Pr(a_{ui}=1 \, | \, \theta_u, \theta_i) = \Pr(\text{guessing} \, | \, n_{\text{opt}}) + \frac{1 - \Pr(\text{guessing} \, | \, n_{\text{opt}})}{1 + e^{-(\theta_{u} - \theta_i)}}
\end{displaymath}

Within the BNE question pool, questions are divided into two main types: single- and multiple-answer questions.
For single-selection questions (unique answer questions), \(P(\text{guessing} | n_{\text{opt}})\) is straightforward, equating to $1/n_{\text{opt}}$.
For multiple choice questions, \(P(\text{guessing} | n_{\text{opt}})\) is the inverse of the sum of the possible ways to select any number of answers $k$ from the available options:
\begin{displaymath}
\Pr(\text{guessing} \, | \, n_{\text{opt}}) = \frac1{
\sum_{k=1}^{n_{\text{opt}}} \binom{n_{\text{opt}}}{k}
}
\end{displaymath}

\subsubsection{Decreasing Uncertainty}

The dynamics of the Elo rating system in educational settings are crucial for accurately assessing the skills and abilities of students and questions. The challenge lies in managing evolving uncertainties, which are inherently dynamic. When new students or questions are introduced to the platform, our information on their true abilities or difficulties is limited, meaning high uncertainty. Consequently, during this initial phase, it is essential for the model to make significant updates to its estimates. As more data accumulates, students engage in multiple attempts, and questions are extensively attempted by a set of students, and the model should naturally become more certain about its estimation of ability levels or difficulty levels. In such cases, the model should reduce the update parameter as confidence in the estimates grows.

In order to meet this challenge, recent applications of the Elo rating system in educational contexts \cite{abdi2019multivariate} have replaced the fixed constant $K$ in Equation~\ref{elo-update-rule} with a dynamic uncertainty function. This function, denoted as $U(n)$, is defined as:
\begin{displaymath}
\mathit{U}(n) = \frac{a}{1 + bn}
\end{displaymath}

\noindent
where \(a\) is the constant hyper-parameter determining the starting value;
\(b\) is the constant hyper-parameter determining the slope of changes;
\(n\) is the number of prior attempts of the user or question parameter.

The exact parameter values, as highlighted by \cite{pelanek2016applications}, carry relatively less weight, as different choices for $a$ and $b$ tend to yield remarkably similar outcomes. In our case, we set $a = 1$ and $b = 0.5$ for both question and user attempts. These values were determined through an optimization process using grid search. However, it is important to mention that our model consistently delivered stable performance, and the precise choice of parameter values had only a negligible effect on the results.

Moreover, in keeping with the central aim of learner models, which is to effectively track shifts in user abilities, we have introduced a lower bound for the uncertainty function applied to user ability. By incorporating this lower bound of $0.03$ into the user uncertainty function, we ensured that user ability updates persist even after a considerable number of attempts. With our current values of $a$ and $b$, this lower bound applies after 65 attempts.

\subsubsection{Multi-tag Knowledge Component Extension}

As previously described, in our BNE data one question may be tagged by multiple specialties. To account for the ability of users in each of the tagged medical specialties separately we used the multi-concept extended version of the Elo rating introduced by \cite{abdi2019multivariate}. The difference is that, instead of having only one global user ability parameter \(\theta_{u}\), we estimated user ability $\theta_{us}$ for each specialty $s$.
It is important to note that, given the absence of information regarding the relative importance of tagged specialties for each question in the data, we adopted a straightforward approach as outlined in \cite{abdi2019multivariate}. Specifically, we computed the mean ability \(\lambda_{ui}\) of student $u$ on question $i$ by averaging user $u$'s abilities across all medical specialties \(s_1, \ldots, s_\delta\) tagging question \(i\), assigning equal weight to each specialty in this calculation.
\begin{displaymath}
    \lambda_{ui} = \frac1\delta \sum_{k=1}^{\delta} \theta_{us_k}
\end{displaymath}
Furthermore, not all specialties may have the same average difficulty level. In order to alleviate this, we define and estimate distinct parameters for question difficulty (\(d_i\)) and specialty difficulty (\(\theta_{s}\)). We denote by $\mu_i$ the sum of question difficulty $d_i$ and the average of difficulties of all skills $s_1, \ldots, s_\delta$ involved in question $i$: 
\begin{displaymath}
    \mu_{i} = d_i + \frac1\delta \sum_{k = 1}^\delta \theta_{s_k}
\end{displaymath}


Thus the probability of answering correctly becomes:
\begin{displaymath}
\ \Pr(a_{ui}=1 \sep \lambda_{ui}, \mu_{i}) = p(\lambda_{ui}, \mu_i) \triangleq \sigma({\lambda}_{ui} - \mu_i) 
\end{displaymath}

The update of the question difficulty $d_i$ remains the same. However, now the update on the user skill parameters $\theta_{us}$ occurs on each tagged specialty separately, and the update for specialty difficulty $\theta_s$ follows a similar pattern as the updates for item difficulty:
\begin{displaymath}
\begin{aligned}
\ d_i := d_i + U(n) \left( p(\lambda_{ui}, \mu_i) - a_{ui} \right)\\
\ \theta_{us} := \theta_{us} + U(n) \left( a_{ui} - p(\theta_{us}, d_i + \theta_s) \right)\\
\ \theta_s := \theta_s + U(n) \left( p(\theta_{us}, d_i + \theta_s) - a_{ui} \right). 
\end{aligned}
\end{displaymath}






It's important to note that while updating \(\theta_{us} \) and \( \theta_{s} \), the prediction formula operates at the specialty level for each tagged specialty, just like in \cite{abdi2019multivariate}, although the \(  d_{i} \) update is based on question-level prediction.

By utilizing the Elo rating system along with the aforementioned extensions, it becomes possible to estimate three critical aspects: user ability in each specialty, questions' individual difficulty, and specialties' global difficulty. 

\subsection{Data Preparation Process}

Before starting to train the Elo rating model and Logistic regression over the 2020-2021 data set, we performed a series of pre-processing steps on the combined data from the training period data set and the mock ECN final exam data set. These steps were carried out in the following order:

\begin{itemize}
    \item Removal of duplicated rows: 267 rows out of 29,900,533 were removed.
    \item Exclusion of questions without any tagged medical specialty: None removed (the data extraction process was already limited to questions with tagged specialties), but $30\%$ lacked specialty tags initially.
    \item Exclusion of questions that are neither unique nor multiple-choice questions (open-answer questions): None removed.
    \item Binarization of question ratings (BNE has a more sophisticated rating scheme depending on the number of correct and incorrect answers ticked).
    \item Removal of users with fewer than 100 interactions: No questions or students were removed during this step. Since all students in the dataset had taken the ECN mock exam, they all had at least 100 attempts.
    \item Removal of questions with fewer than 100 interactions: $79.11\%$ of the unique questions were removed.
\end{itemize}

As a result, our training period data set now consists of 22,294,780 attempts, made by 8,616 distinct users to 74,704 unique questions across 31 medical specialties. The mock ECN final exam data set includes 3,172,546 attempts. Within that data set, there are 372 unique questions taken by 8,616 users across 28 distinct medical specialties.

 \begin{figure*}
  \centering 
  \includegraphics[width=0.8\linewidth]{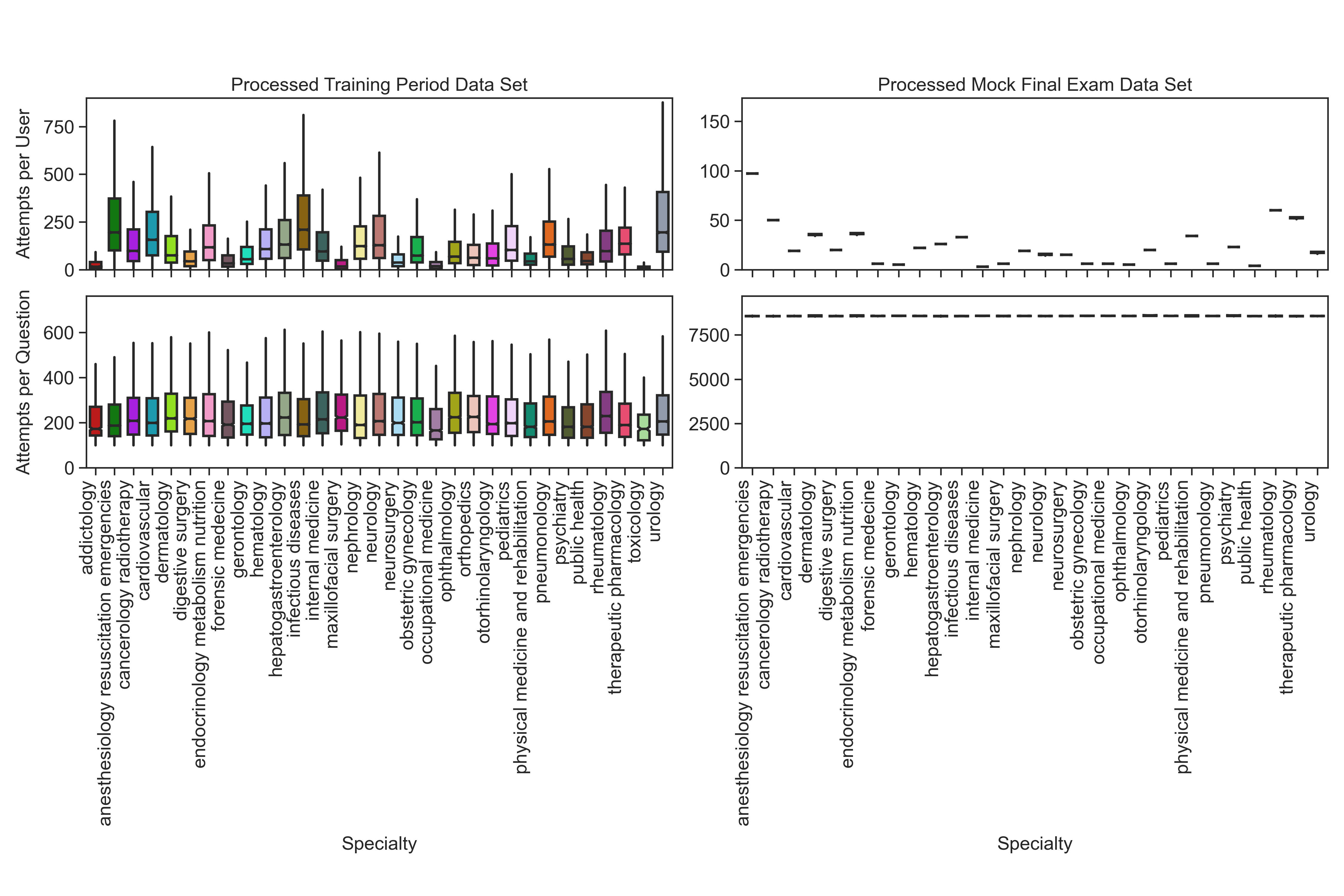}
  \caption{Number of Attempts by Each User and to Each Question across the 31 Medical Specialties. \\
The top left box plot shows the distributions of the number of attempts by each user across the 31 specialties. The bottom left box plot depicts the number of attempts to questions in each specialty.
During the mock exam, all students took identical questions, resulting in quasi-uniform numbers of attempts given by students and received by questions (top and bottom right plots).}
  \Description{}
      \label{fig:attemptplot}
\end{figure*}

Figure~\ref{fig:attemptplot} offers a visual depiction of the distribution of answers across students, questions, and specialties in both the training period and the mock final exam data sets after the pre-processing. 

\subsection{Information Encoding \& Initialization of Elo Ratings via Logistic Regression Outputs}

As previously mentioned, in the Elo rating system, initial estimates for both questions and users are typically set to 0, which can lead to high uncertainty. To address this, an alternative approach is to use the logistic regression outcomes of the previous year's data as informed initial values for initializing Elo ratings, rather than starting from scratch. With this approach, the Elo rating algorithm is anticipated to converge faster and more accurately, providing a "head start" that conserves computational resources and leads to more precise estimates.

To prepare the extensive BNE dataset for logistic regression modeling, we employed a one-hot vector encoding method inspired by \cite{vie2019knowledge}. This technique transformed each attempt in the original data set into a sparse vector containing all relevant information. In our adaptation of this approach, we aimed to closely align our logistic regression model with the principles of Elo rating estimations, while also incorporating all the aforementioned extensions we applied for the Elo rating system. To achieve this, we included the one-hot encoding of user-specialties interaction, question, and specialty for each attempt. With this approach, the logistic regression was able to estimate users' ability in each specialty, the difficulty of individual questions, and the overall difficulty of each specialty.

To implement the initialization approach, an essential step involves comparing the logistic regression model against the Elo rating system, utilizing data from the same year. This step aimed to ensure that, before employing the logistic regression model on the previous year's data and utilizing its outcomes for initialization purposes, the model aligned with the Elo framework, generating compatible and consistent results.

Subsequently, we applied the logistic regression model to the data from the 2019-2020 educational year, utilizing the outcome estimates as initial values for the Elo rating process applied to the 2020-2021 data. Our data set for the academic year 2019-2020 (spanning from September 15, 2019, to March 1, 2020) includes 400,774 distinct questions and 25,978 unique users. However, after filtering data to retain questions and users with enough attempts (cf. above), only 47,579 questions and 8,239 users were shared between 2019-2020 and 2020-2021.

As a result, we initialized the question difficulty and student ability values for the 2020-2021 academic year using the estimates obtained from the logistic regression model applied to the 2019-2020 data, whenever these values were available. In cases where values were not present in the 2019-2020 data set for a particular question or student, we initialized their 2020-2021 values to zero. 

In order to allow the uncertainty function to be able to further update those initialized values, without entirely destabilizing the estimations, we set the initial number of previous attempts to 50, which seemed a reasonable compromise between the actual number of attempts (>100 which would make any update negligible) and 0 (which would underweight the previous history). 

\subsection{Performance Evaluation Metrics}

In our performance evaluation, we examined the effectiveness of two variants of the Elo rating system on the training data set. One variant initialized all ability and difficulty values to 0, while the other initialized values based on logistic regression from the previous year. We also compared these Elo variants with logistic regression on the entire training data set. We used several key statistics, including Area Under the Curve (AUC), Root Mean Squared Error (RMSE), and Accuracy (ACC), to assess the prediction performance of these models first on the training period and second on the mock exam. 

First, to comprehensively evaluate the prediction capabilities of the Elo rating system throughout its iterations, mirroring its real-world use within the platform, and understand the impact of initializing estimates via logistic regression, we compared these three models during the training period. We calculated the AUC, ACC, and RMSE scores for each training period day from September 16, 2020, to March 14, 2021, providing insights into how these models adapted and remained robust over time. 

Subsequently, we turned our attention to evaluating these models' ability to predict performance on the mock exam using the same metrics: AUC, RMSE, and ACC. However, the mock exam posed a unique challenge as it featured entirely new questions that were not part of the training period. To address this challenge, we needed to estimate the difficulty of these mock exam questions. Our approach involved combining the entire training data set with a random selection of 60\% of user data from the mock exam data set as the train set while designating the remaining 40\% of user data from the mock exam data set as the test set. This allowed us to create a training set that encompassed all attempts, including those from the mock exam, for 60\% of users. For the remaining 40\% of users, we included only their attempts from the training period in the train set. By doing so, when we ran the learner models on the training set, we obtained estimates of ability for all students and difficulty for all questions, which in turn enabled us to measure the models' prediction ability on the mock exam.

Following this data division process, the training subset comprised a substantial 24,152,933 entries, involving 8,616 unique users and 74,971 unique questions. In parallel, the test subset consisted of 1,268,752 entries, encompassing 3,447 unique users and 372 unique questions.

To assess the prediction ability of the models on the mock exam, after executing the models on the specified training set and obtaining difficulty estimates for all questions and ability estimates for all students, we evaluated its prediction performance on the test set. This evaluation capitalized on the stabilized estimates derived from the comprehensive training data set.

\begin{figure}
  \centering
    \includegraphics[width=1\linewidth]{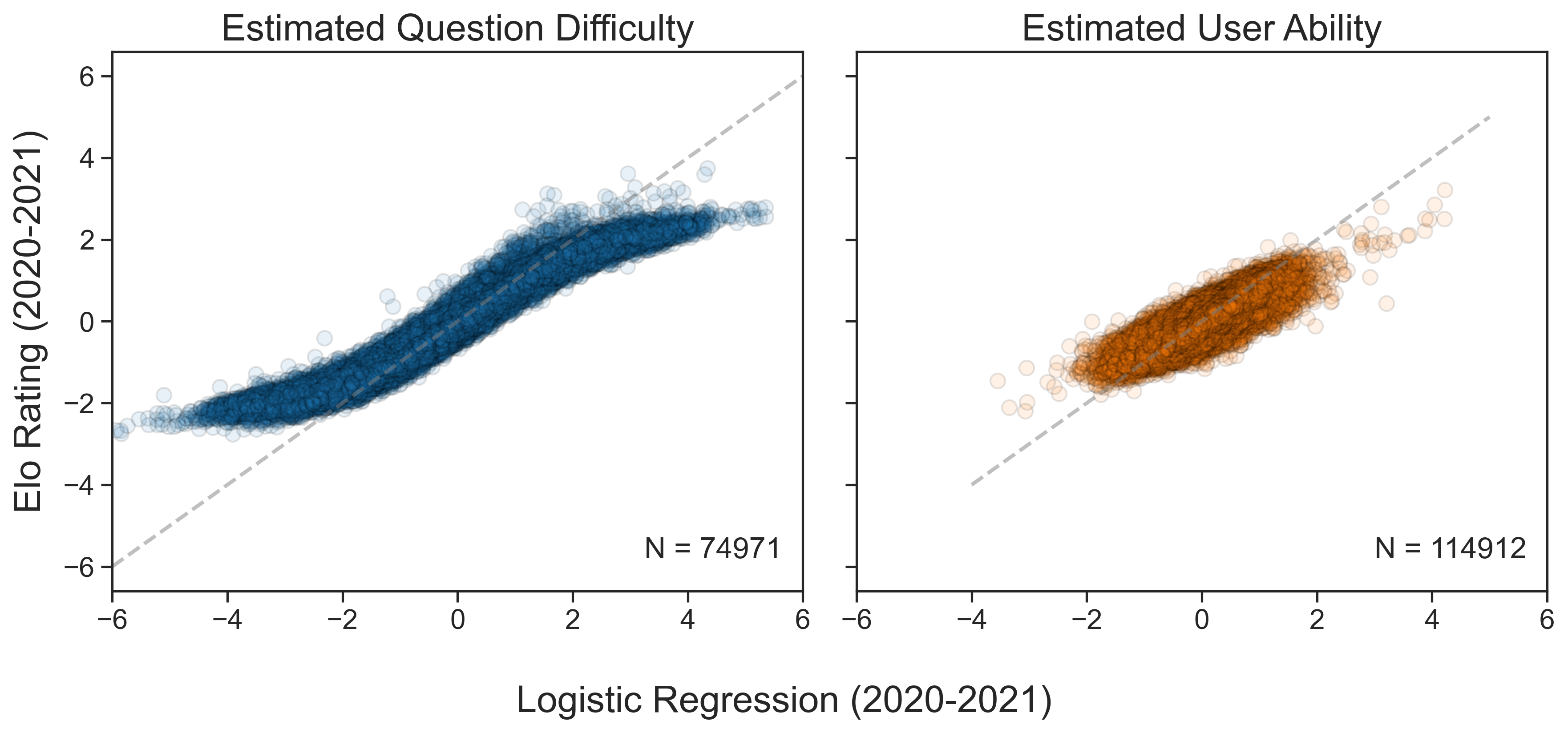}
    \caption{Comparing Logistic Regression and Elo Rating outcomes for question difficulty (left) and user ability across 31 specialties (right) in the same 2020-2021 dataset. Scatter plots illustrate the alignment, with $y=x$ lines for reference. The left plot displays Logistic Regression difficulty estimates on the $x$-axis and Elo Rating estimates on the $y$-axis. On the right, the plot contrasts user ability estimates, with Logistic Regression on the $x$-axis and Elo Rating on the $y$-axis. Sample sizes ($N$) are included in each plot.}
  \label{fig:comparison_logreg_elo}
\end{figure}

\section{Results}

\subsection{Correlation between the Estimates}

Figure~\ref{fig:comparison_logreg_elo} illustrates a notable positive correlation between the estimates of question difficulty ($r=0.97$) and user ability ($r=0.86$) derived from the Elo rating and logistic regression models for the same-year data. This strong positive correlation clearly indicates that both models converged toward similar final estimations regarding question difficulty and user ability.

\begin{figure}
  \centering
    \includegraphics[width=1\linewidth]{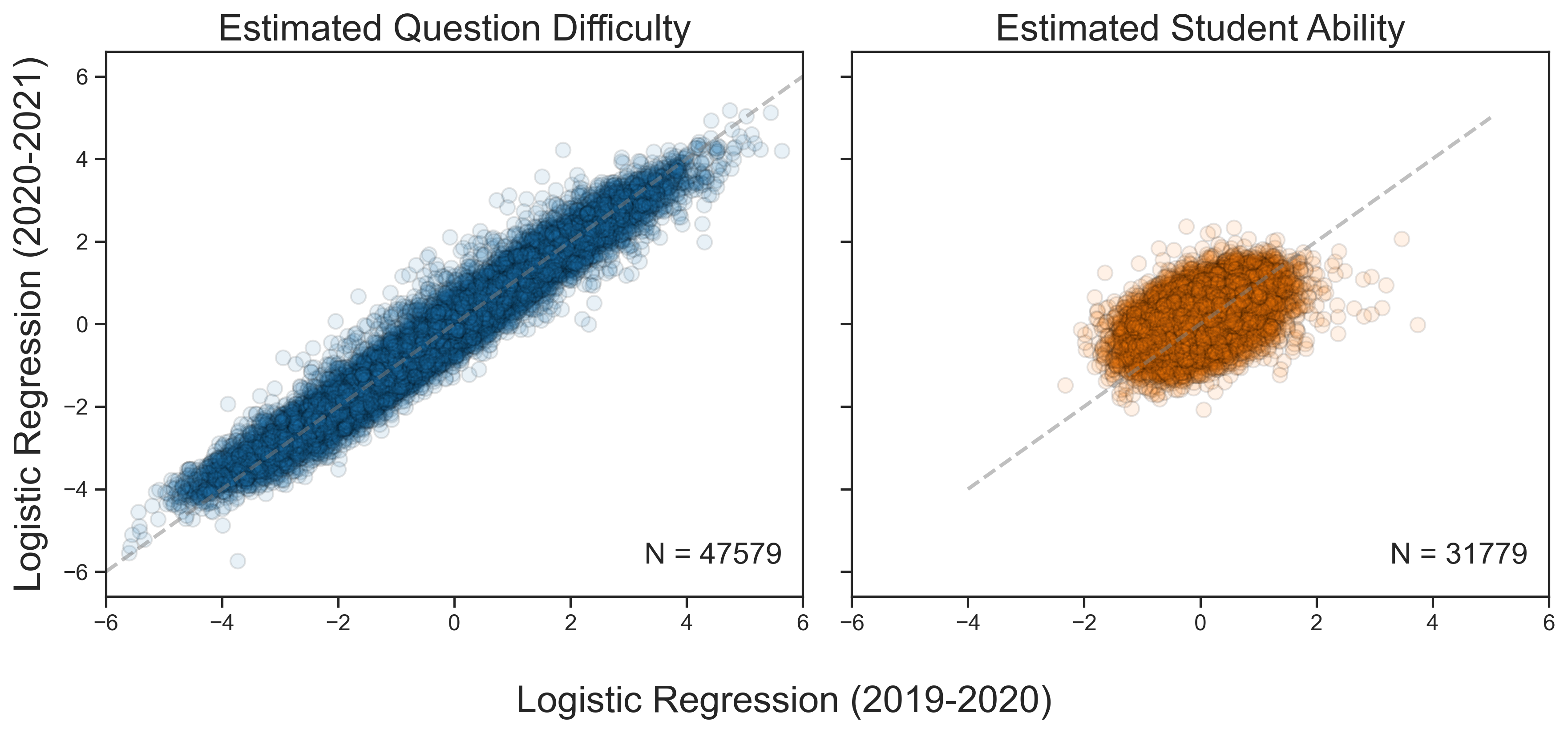}
    \caption{Comparing Logistic Regression Outcomes. \\
    Estimated question difficulty (left) and user ability on each of 31 specialties (right) in the two successive education years (2019-2020 and 2020-2021) using the Logistic Regression model. $y=x$ lines are given for reference.
    }
          \label{fig:comparison_old_new_logreg}
      \end{figure}

Figure~\ref{fig:comparison_old_new_logreg} shows the question difficulty and student ability estimations using the logistic regression in the two successive years. Specifically, for shared questions, we observe a robust positive correlation of 0.98, indicating that the difficulty levels of these questions remain relatively consistent over time. However, when it comes to students' abilities, the correlation, albeit positive at 0.54, is not as strong. This suggests that students' abilities in various specialties have undergone some changes over the years, as we anticipated.
      
\subsection{Prediction Accuracy}

\begin{figure*}
  \centering
  \includegraphics[width=0.6\linewidth]{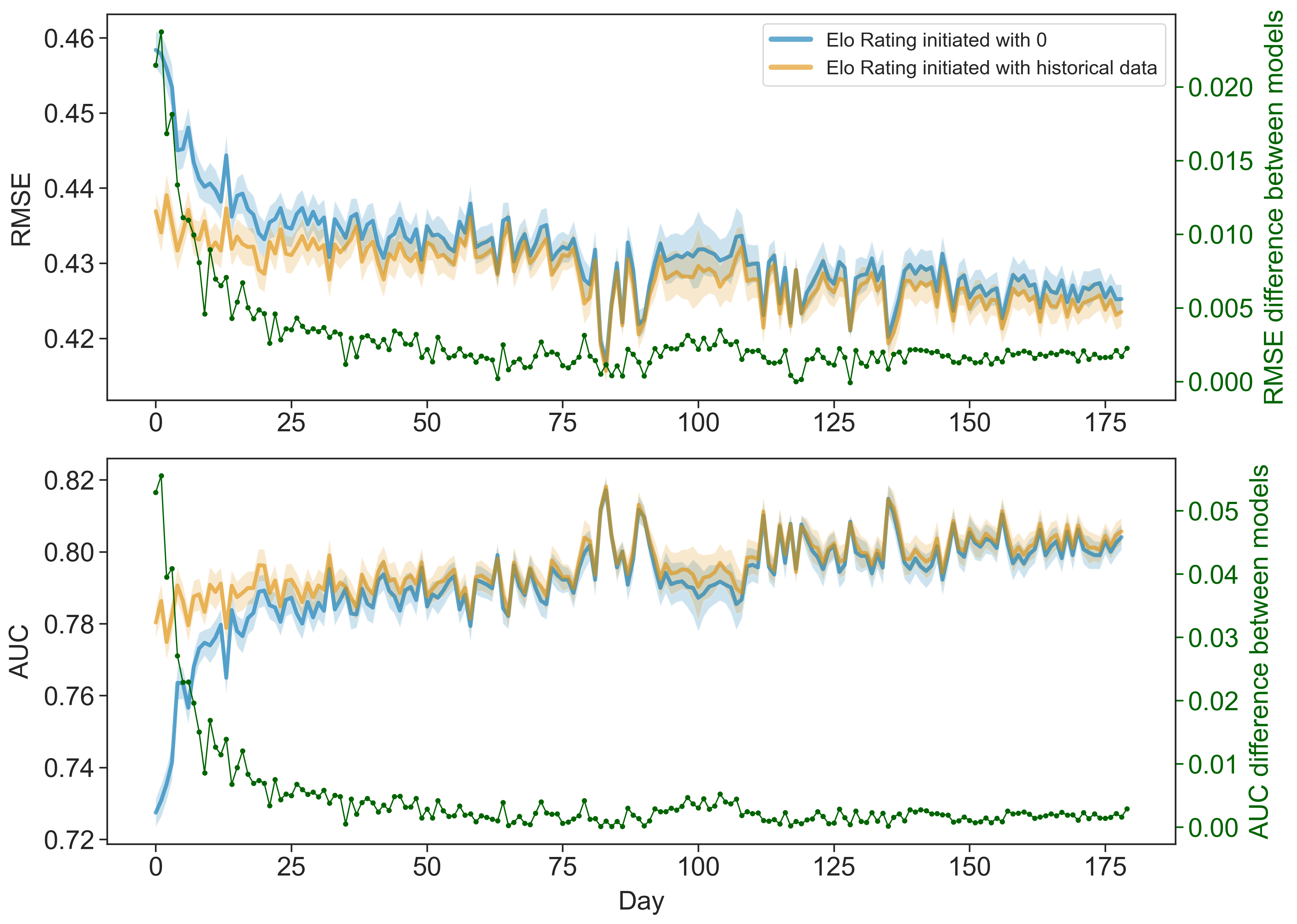}
  \caption{Comparing Models' Prediction Performance During Training.\\
  RMSE (top) and AUC (bottom) values as a function of training days, for two versions of the Elo rating system. Shaded regions around the mean lines represent the 95\% confidence intervals calculated using the standard error. A secondary $y$-axis on the right side illustrates the absolute difference between the two models with the green line plot.}
  \Description{}
\label{fig:rmse_auc_acc_evolution}
\end{figure*}

Figure~\ref{fig:rmse_auc_acc_evolution} presents a visual representation of the RMSE and AUC scores over 180 consecutive training days for each model. Findings indicate a substantial prediction accuracy advantage at the beginning of the training year when initializing the ability and difficulty values based on the data from the previous year. This advantage is reflected in an initial boost in average accuracy (+0.016 AUC) and a reduction in the average error (-0.008 RMSE) during the initial 30 days of training. However, it's worth noting that the initial disparity between the two model versions diminishes rapidly and becomes less than 1 point within a few days. By the end of the training period, the advantage of initializing with historical data becomes nearly negligible, with only a marginal improvement in average accuracy (+0.002 AUC) and a minimal reduction in average error (-0.002 RMSE) observed during the last 30 days of training.

\begin{table*}
  \caption{Comparing Models' Prediction Performance on Mock Exam}
  \label{tab:resultsummary}
  \begin{tabular}{ccccccc}  
    \toprule
    Model & RMSE ($\downarrow$) & AUC ($\uparrow$) & ACC ($\uparrow$) & Precision ($\uparrow$) & Recall ($\uparrow$) & F1 ($\uparrow$) \\
    \midrule
    Elo rating initialized at 0 & 0.419 & 0.812 & 0.737 &  0.717 & 0.685 & 0.701\\
    Elo rating initialized with historical data & 0.418 & 0.813 & 0.738 & 0.720 &  0.684 & 0.702\\
    Logistic regression & 0.419 & 0.811 & 0.736 & 0.714 & 0.689 & 0.701 \\
    \bottomrule
  \end{tabular}
\end{table*}

Table~\ref{tab:resultsummary} shows the prediction performance on the mock exam for the three models: Elo rating initialized at 0, Elo rating initialized historical data, and logistic regression. These results reveal that the three models show highly similar prediction accuracy, with a slightly better performance on the Elo rating model initiated with historical data over other models.


\section{Discussion}

This research is an initial step in integrating the multi-concept Elo rating system into our medical training platform in order to achieve real-time estimates of user performance. The results demonstrated the Elo rating system's comparable prediction power to the logistic regression models, confirming its suitability for this specific data set.

The Elo rating system offers several significant advantages in the context of our medical training platform. Firstly, the multi-concept Elo rating system excels in estimating concept-level competencies, which is crucial for tailoring adaptive learning experiences in our data in which questions mostly require knowledge from multiple medical specialties. Secondly, it stands out as a computationally efficient and cost-effective option especially in real-time estimations, compared to logistic regression models which require processing a vast amount of previously collected data. 

Given our primary objective of identifying the most effective prediction model for online applications in our platform, and considering the challenges associated with logistic regression in terms of online adaptability, our focus naturally shifted towards a more detailed comparison of the two versions of the Elo rating system: one starting from scratch at the beginning of the year, and one with difficulty and ability values initialized based on the previous year's data. 

Our findings underscore that while the overall performance in predicting mock exam results remains very similar regardless of the initialization approach, a distinct advantage emerges for initialization based on historical data, particularly during the initial phases of iteration. This may be important in scenarios that demand accurate estimations from the outset, such as real-time or online applications. This approach of initializing the model with historical data enhances the model's ability to produce quicker and more precise estimates, thereby enhancing the reliability of personalized learning environments utilizing Elo rating systems.

\subsection{Unique Characteristics of the Data set}

Our data set has very broad knowledge components, made of 31 distinct medical specialties, each of which is a huge corpus of information. Moreover, this platform prioritizes comprehension over memorization: questions are hardly ever repeated twice. Students have to generalize their knowledge while attempting the questions. This departure from purely memory-based learning provides an excellent chance to assess the model's performance in dealing with non-repeated inputs. While this limited exposure to questions challenges standard prediction models used for repeated question iterations, it also allows us to evaluate the model's flexibility in settings where students rely on wider conceptual knowledge rather than memorized responses. Additionally, our platform allows students to personalize their own training experiences. They have the option of selecting the medical specialty in which they want to train and the type of questions, resulting in unique interaction patterns for each student. This adds another layer of complexity to our data set. In addition to these complexities, we also lacked control over learning occurring outside the platform. 

Despite these challenges, it is remarkable that the Elo rating system has achieved significant prediction power for assessing the accuracy of students' future responses, with about $73.7\%$ accuracy and $0.81$ AUC. This demonstrates the model's adaptability and endurance in circumstances that deviate from standard educational data. In addition, the Elo rating system has shown good online prediction accuracy during training, right from the start when initializing with historical data, and after about 15-20 days when starting from scratch.

With the overarching goal of transforming our medical training platform into a personalized and adaptive format through knowledge tracing methods, we carefully considered the data set's characteristics. The Elo rating model stood out as a prime choice due to its simplicity, rapid parameter estimation, and real-time knowledge assessment capabilities. Its straightforwardness, coupled with widespread use in applications like online games and chess, makes it easily explainable compared to more complex models. An exemplar of transparency in a multivariate Elo version is evident in the study by \cite{abdi2019multivariate}. The study demonstrates the feasibility of making the algorithm transparent to students, a practice that not only heightened their motivation to engage with the platform but also enhanced their trust in the recommendations provided.

The implemented extensions further enhance adaptability to our data set's unique characteristics, offering optimization along with advantages in suitability and transparency. Notably, the multi-concept Elo rating model, in contrast to its single-concept version, acknowledging the non-transitive nature of skills, provides a realistic representation of learners' capabilities, crucial for accommodating interdependencies within medical specialties, potentially involving prerequisites. Thus, the multi-concept Elo rating model emerges as a fitting and transparent knowledge-tracing method for our complex data set. 

\subsection{Limitations and Further Work}
One major limitation is that our tested models (logistic regression and Elo rating) do not consider the natural forgetting of knowledge over time, which is well-documented in human memory research dating back to Ebbinghaus in 1885 \cite{ebbinghaus2013memory}. Incorporating learning and forgetting curves into prediction models, as shown in research like DAS3H \cite{choffin2019das3h} and MV-Glicko \cite{abdi2021modelling}, which builds upon the multivariate-Elo rating system \cite{abdi2019multivariate}, can improve the models' prediction accuracy. 

Our study also suggests several promising future research directions. One area of focus is improving learning models to better assess question difficulty and student ability, especially for topics requiring knowledge of multiple concepts. This involves determining the importance of each concept in problem-solving and investigating how these concepts interact during the learning process. Additionally, comparing the Elo rating system with the Glicko rating model within our data set could provide insights into the role of learning and forgetting curves in student performance estimation.

Beyond the aforementioned research areas, a critical future direction involves implementing the Elo rating system for online recommendations regarding specialties and question difficulty. However, this endeavor presents challenges in platforms where questions often have multiple specialty tags. As underscored by the research findings of \cite{cepeda2008spacing}, the premature revisiting of knowledge can exert detrimental effects on long-term memory. Consequently, during the recommendation phase, it becomes imperative to not only select questions aligned with the student's current needs but also to ensure that these questions do not encompass specialties that may not be relevant to the student's current stage of learning. To address this, we can consider a new approach that calculates students' abilities based on combinations of specialties, rather than individual ones. This new strategy could substantially enhance the model's efficacy when suggesting questions that align with students' learning requirements and are pertinent to their current learning stage. Such an approach would ensure that students are presented with a tailored set of questions that optimally support their progress while avoiding the unnecessary revisiting of topics that might hinder long-term retention—a crucial consideration for the success of an adaptive learning platform.

\section{Conclusion}
In conclusion, our study underscores the remarkable adaptability of the Elo rating system to the intricate challenges posed by a large, sparse, and multifaceted data set, where questions are tagged with multiple knowledge components. The Elo rating system, along with its enhanced version that leverages historical data for initial estimations, has exhibited a commendable level of prediction accuracy. 

These results offer reassurance regarding the Elo system's robustness and versatility, emphasizing its capability to provide reasonable predictive value even in complex situations. This insight is crucial for the broader learning analytics community, providing confidence in the effectiveness of the Elo rating system as a predictive model in educational settings. Overall, the study contributes to the ongoing discourse on learning analytics methodologies, offering practical insights and encouraging further exploration of the Elo rating system's applicability in diverse learning scenarios.

\begin{acks}
This research was funded by Agence Nationale de la Recherche, grants ANR-17-EURE-0017, ANR-10-IDEX-0001-02 and ANR-21-CE28-0030.
\end{acks}
\bibliographystyle{ACM-Reference-Format}
\bibliography{myref}

\end{document}